\documentclass[usenatbib]{mnras}

\usepackage[english]{babel}
\usepackage[utf8]{inputenc}
\usepackage[T1]{fontenc}
\usepackage{ae,aecompl}

\usepackage{amsmath}
\usepackage{dsfont}
\usepackage{amssymb}

\usepackage{graphicx}
\usepackage{natbib}
\usepackage{url}

\usepackage{tikz}
\usetikzlibrary{arrows,shapes}

\usepackage{rotating}
\usepackage{xcolor}

\usepackage{listings}

\usepackage{hyperref}
\usepackage{url}

\newcommand{\vect}[1]{\boldsymbol{#1}}

\newcommand{\comment}[1]{\textcolor{black}{#1}}

\title{Numerical aspects of Giant Impact simulations}
\author[C. Reinhardt and J. Stadel]{Christian Reinhardt,$^1$\thanks{christian.reinhardt@ics.uzh.ch}
									Joachim Stadel$^2$\\
									Institute for Computational Science, University of Zurich, 8057 Zurich, CH
									}

\date{Accepted XXX. Received YYY; in original form ZZZ}

\pubyear{2016}

\begin{document}
\label{firstpage}
\pagerange{\pageref{firstpage}--\pageref{lastpage}}
\maketitle

\begin{abstract}
In this paper we present solutions to three short comings of Smoothed Particles Hydrodynamics (SPH) encountered in previous work when applying it to Giant Impacts. First we introduce a novel method to obtain accurate SPH representations of a planet's equilibrium initial conditions based on equal area tessellations of the sphere. This allows one to imprint an arbitrary density and internal energy profile with very low noise which substantially reduces computation because these models require no relaxation prior to use. As a consequence one can significantly increase the resolution and more flexibly change the initial bodies to explore larger parts of the impact parameter space in simulations. The second issue addressed is the proper treatment of the matter/vacuum boundary at a planet's surface with a modified SPH density estimator that properly calculates the density stabilizing the models and avoiding an artificially low density atmosphere prior to impact. Further we present a novel SPH scheme that simultaneously conserves both energy and entropy for an arbitrary equation of state. This prevents loss of entropy during the simulation and further assures that the material does not evolve into unphysical states. Application of these modifications to impact simulations for different resolutions up to $6.4 \cdot 10^6$ particles show a general agreement with prior result. However, we observe resolution dependent differences in the evolution and composition of post collision ejecta. This strongly suggests that the use of more sophisticated equations of state also demands a large number of particles in such simulations.

\end{abstract}


\section{Introduction}
Collisions are one of the most fundamental processes in planet formation as they are a key mechanism to explain growth of initially micron sized dust grains to massive planets like the ones we observe today in our solar system. The last stage of terrestrial planet formation is dominated by very energetic collisions between roughly Mars-sized planetary embryos called Giant Impacts (GI) \citep{Agnor1999}. Such GI play a key role in the process of planet formation as they determine and influence many important properties of the final planetary system like the final number, mass and spin state \citep{Agnor1999}, chemical and physical composition \citep{Benz2007, Wiechert2001, Genda2005, Tonks1992, Jutzi2013} and the presence of satellites (e.g. \citet{Benz1986}, \citet{Canup2004}, \citet{Citron2015}).

Starting with Benz and Cameron's \citep{Benz1986} pioneering work in the mid 80's simulations of GI between planetary embryos have became an increasingly important tool to study the outcomes of such violent collisions. While most of the earlier work (e.g. \citet{Benz1987}, \citet{Cameron1991}, \citet{Canup2001}, \citet{Canup2004}, \citet{Wada2006}) concentrated on the giant impact hypothesis of the origin of the moon \citep{Cameron1976} more recent simulations explore a variety of possible scenarios. Some recent examples include satellite formation in a more general sense (e.g. \citet{Canup2011}, \citet{Citron2015}), mantle stripping events \citep{Benz2007} or deriving merging criteria to improve N-Body simulations \citep{Genda2012, Leinhardt2012}. 

Most prior work uses the Smoothed Particles Hydrodynamics (SPH) method to solve the equations of motion of the material. Since SPH interpolates the physical quantities using a discrete set of particles that evolve with the flow, it can simulate very deformed geometries and track the ejected material over a large dynamic range making it an ideal tool for such simulations. For the research presented in this paper we modified the cosmological hydrodynamics code GASOLINE \citep{Wadsley2004} that has been extensively used in many astrophysical applications and applied it to such GI simulations. The code uses a modern SPH implementation, a tree to calculate gravity and has a modular design that allows one to implement new physics without touching the complicated parallel layer. We have implemented the Tillotson equation of state (EOS) \citep{Tillotson1962} within GASOLINE to model the condensed materials that planets are made of. To facilitate the use of the Tillotson EOS we developed a library\footnote{The TillotsonEOS library is freely available under \url{https://github.com/chreinhardt/tillotson.git}. We also provide our code to create low noise particle representations of planets, \texttt{ballic}, at \url{https://github.com/chreinhardt/ballic.git}.} that provides functions to calculate the pressure, sound speed and temperature allowing its use in any simulation software. Besides the SPH modifications described in this paper we also added a new class of smoothing kernels \citep{Dehnen2012} that do not suffer from particle clumping and allow increasing the number of neighbours thereby reducing the noise of the method.

In this paper we present solutions to three short comings of SPH that previous work encountered. First of all it is very difficult to get an accurate SPH representation of a planet's equilibrium initial condition (IC) so one has to evolve ("relax") the system carefully to reach a true equilibrium state before doing any impacts. This is very time consuming and can, in the worst case, use more computational resources than the actual simulations. In Section~\ref{section:ic} we present a new method to generate extremely low noise planetary models that do not need relaxation prior to a collision which drastically reduces the total simulation time.

Another issue is the proper treatment of the material--vacuum boundary at a planet's surface, where standard SPH tends to severely underestimate the material's density causing it to jump from the condensed to an expanded state. This creates an "atmosphere" of low density material around a planet prior to the impact. A solution to this problem is discussed in Section~\ref{section:newdens} where we present a modification of the standard SPH density estimator that properly calculates the density for material--vacuum boundaries encountered at the surface of the planets. Another possible remedy would be to apply DISPH \citep{Saitoh2013} to GI simulations as done by \citet{Hosono2016}. Because this SPH scheme does not use the density but pressure as a smoothed (and thus differentiable) quantity, it performs well at contact discontinuities and free surfaces, e.g., encountered at the core-mantle boundary and the planet's surface. In the present work we focus on undifferentiated bodies. However, the proper treatment of the core--mantle boundary is not a problem and the improvements described here have already seen application in realistic models with an iron core and basalt mantle. 

Finally, SPH does not necessarily conserve both energy and entropy for an arbitrary equation of state. In the case of an ideal gas one can define an entropy function \citep{Springel2002} which allows one to conserve entropy for purely adiabatic flows. For a more complex, maybe even tabular EOS, this is in general not possible. In Section~\ref{section:isph} we present a new, entropy conserving SPH scheme that can be applied to any equation of state. This prevents loss of entropy during the simulation and assures that the material does not evolve to unphysical states thus also eliminating code crashes resulting from particles reaching negative internal energies. Code tests and applications of the methods to GI simulations are presented in Section~\ref{section:tests}. Finally a summary of the present work and conclusions are discussed in Section~\ref{section:conclusions}.

\section{Creating quiet particle realizations of planets}
\label{section:ic}

In Smoothed Particles Hydrodynamics (SPH) the density of each particle (as well as pressure forces and changes in internal energy) is derived from the positions of its neighbours. Thus it is crucial to have an accurate particle representation of the initial conditions. An inhomogeneous distribution will cause large spurious radial and angular density fluctuations that cause oscillations and requires careful relaxation. Having Poison noise in the IC not only means that we waste computational time on relaxing the models but is in general problematic for SPH as it affects the method's convergence \citep{Monaghan1985}. For larger simulations with several million particles, relaxation can take a substantial part of the total simulation time and doing such simulations becomes prohibitive if the IC is too noisy. This makes the traditional approach laborious for parameter studies where the target and impactor properties are varied. 

Previous work placed the particles on a uniform 3D lattice either with constant internal energy \citep{Genda2012} or e.g., an isentropic thermal profile \citep{Canup2012}, inside a sphere to have a very uniform and thus low noise particle distribution. To obtain a density gradient that is consistent with the thermal profile and get rid of the Poison noise at the boundary, these models are then evolved with the hydrodynamics code (usually with a velocity damper that reduces the particles velocities by a given fraction after each step) until an equilibrium representation is reached. This method has two short comings: 1) it does not allow one to imprint a (radial) density gradient and 2) also has severe noise at the planet's surface where the grid is not adapted to the spherical symmetry of the profile. Some authors have tried to model a density gradient by distorting the uniform grid radially \citep{Woolfson2007} but the resulting models were out of equilibrium and still needed relaxation.

In the present paper we suggest a different approach that respects the spherical symmetry of the problem and provides a very uniform radial particle distribution by using an equal area tessellation of the sphere. This method not only produces very low radial and angular density fluctuations but it also allows one to imprint a density gradient that closely follows the equilibrium solution thus making relaxation of the models obsolete.

Prior to generating the SPH representation of the planetary bodies one has to (numerically) solve the usual internal structure equations (e.g., \citet{Alibert2014}). For our boundary conditions (e.g., $M \left( r = R \right) = M_{tot}$ and $\rho \left( r = R \right) = \rho_0$) it is convenient to solve these for $\rho \left( r \right)$, $M \left( r \right)$ and $u \left( r \right)$. Besides the equation for hydrostatic equilibrium
\begin{equation}
\frac{\nabla P \left( \rho, u \right)}{\rho} = - G \frac{M \left( r \right)}{r^2}
\label{eqn:hydrostaticequilibrium}
\end{equation}
one needs to specify an equation of state and an internal energy profile to have a closed set of equations. For the present work we used the Tillotson equation of state (EOS) \citep{Tillotson1962} that was originally developed to model hyper-velocity impacts and has been used in many prior simulations (e.g. \citet{Benz1987}, \citet{Canup2001}, \citet{Marinova2011}, \citet{Genda2012} and \citet{Jutzi2013}). Despite the simple analytic form the results are in good agreement with measurements \citep{Benz1986, Brundage2013} and its ability to properly reproduce the materials Hugoniot curve, thus accurately modelling shocks, is excellent \citep{Brundage2013}. More details on the EOS and the material parameters used in this study can be found in Appendix~\ref{appendix:eos}.

The simplest internal energy profile is uniform and each particle has the same internal energy. This works well for bodies smaller than a few hundred kilometres where the density is roughly constant because self--gravity is too weak to significantly compress the material (e.g. \citet{Genda2015}). More massive planets have increasingly steep density gradients, where the use of a uniform thermal profile would be unphysical and furthermore some particles tend to end up below the minimal allowed energy for the material (the so called  cold curve)\footnote{It should be noted that unlike in the case of an ideal gas, isothermal does not mean that the internal energy is constant as there is an additional contribution from the cold curve to the total energy in case of a condensed material EOS (see Appendix~\ref{appendix:eos} for details). Issues with particles falling below the cold curve due to an unphysical, uniform internal energy profile would also readily provide an explanation for the problems with ANEOS reported in \cite{Canup2004}} when starting from a such initial conditions. Such a profile is also not isentropic meaning that the model is expected to exhibit convection until an internal energy profile with constant entropy is reached.

\comment{We imprinted an isentropic thermal profile in our IC, so that the surface temperatures are below 2000 K. This seems sensible in light of the fact that we expect a "cold" surface because the time between GI is typically large enough to allow the planet's surface to cool. Of course, it is possible to solve for higher surface temperature models in the case that the time between collisions is short or the cooling time is long.}

To create the particle realization of an equilibrium model we divided the sphere into concentric spherical shells and use an equal area tessellation of the surface of each shell. We experimented with two tessellation methods, the Icosahedron package \citep{Tegmark1999} and HEALPix \citep{Gorski2005} that was originally developed for analysing data from cosmic microwave background measurements. Both methods in principle suit our purpose. Icosahedral tessellation produces a nearly perfect tangentially homogeneous grid but suffers from larger radial deviations. HEALPix on the other hand produces a better radial distribution but shows a small amount of tangential density fluctuations due to artefacts resulting from the grid. Since, in our case, radial fluctuations are more problematic and the HEALPix grid adapts more flexibly in the inner part of the sphere (see below), it has been used exclusively in all of the work presented here.

We begin particle generation by first solving the above structure equations (Equation~\ref{eqn:hydrostaticequilibrium}). Our method then iteratively minimizes the axis ratios of the finite volumes so that the tangential to radial size \comment{become} almost identical by adjusting the number of particles for each shell and the spacing between shells keeping the particle mass constant. This procedure results in somewhat more or somewhat less particles than specified, meaning that the total mass is deviating from the desired value. We then recalculate the particle mass given the obtained number of particles and repeat the above until we converge. Our final model has the exact total mass and equilibrium density profile (Figure~\ref{fig:densityprofile}) for the planet as well as having equal mass for all particles. There are somewhat larger density fluctuations in the inner--most shells because the number of particles, and thus the radial size of the cells, can not be adapted as flexibly due to constraints from the HEALPix grid. \comment{HEALPix uses 12 ``rectangular'' patches on the sphere each of which can be further subdivided into $n \times n$ pixels. An Icosahedral grid uses 20 ``triangular'' patches, meaning that the inner--most shell has 20 particles which is somewhat more restrictive than the 12 particle shell that is possible with HEALPix.\footnote{We also include a single central particle, although this is optional.}} We also notice that the smoothed density deviates significantly for the imprinted profile in the outer most shells. There the resulting density as determined by the SPH code is lower than that originally imposed, which causes the bodies to be out of equilibrium. For a condensed material EOS this is even more problematic as the estimated surface density is lower than the material's reference density ($\rho_0$, the density of the material at zero temperature and pressure) which causes the material to fall into an unphysical regime. For a colder model with a low surface temperature this is less problematic as the Tillotson EOS in this region has the same analytic form as in the condensed states. For hotter models, however, the particles end up in the intermediate expanded states where the EOS is different and the models become very unstable. It is therefore crucial to correct the surface density as calculated by the SPH code and is made possible by the method we present in the following section.

\begin{figure}
	\centering
	\includegraphics{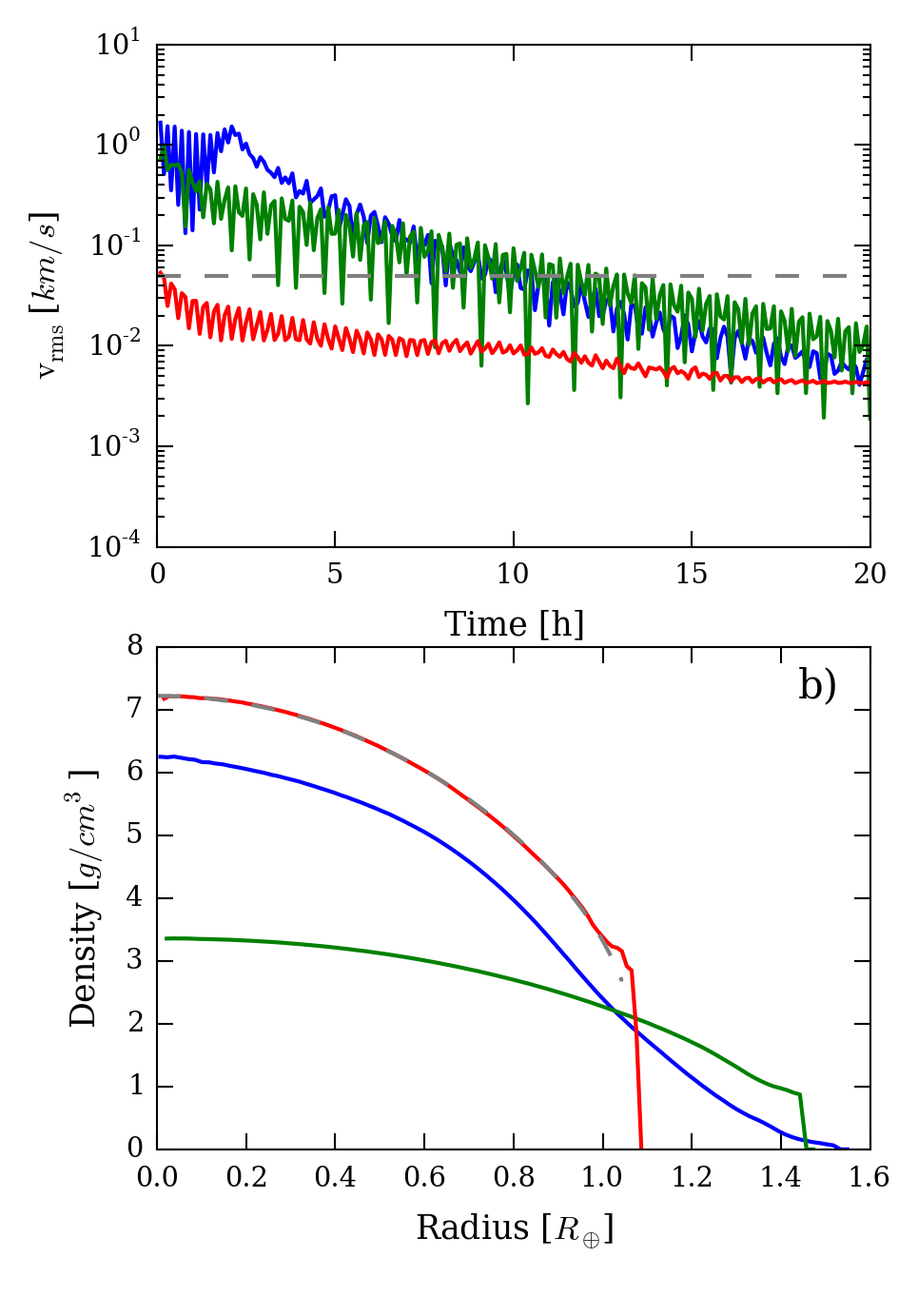}

	\caption{\comment{In this figure we compare \texttt{ballic} (red) to a more conventional method that places particles on a Cartesian grid with a uniform (green) or isentropic internal energy profile (blue). In both cases we evolve the particle representation of an Earth-mass granite model using ISPH (Section \ref{section:isph}) and the density correction proposed in Section \ref{section:newdens}. The top frame (a) shows how the root mean square velocity of the models changes with time. The dashed grey line marks $v_{\rm RMS} = 50 ms^{-1}$, below which models have typically been considered to be relaxed ($v_{\rm RMS}$ < $5 \cdot 10^{-3} v_{\rm imp}$, with an assumed impact velocity of $v_{\rm imp} = 10 ms^{-1}$). The plot clearly show that an initial distribution made with \texttt{ballic} is relaxed right from the start, while the other two models require about 10h in simulation time to reach this point. Even after that they remain noisy and show oscillations in $v_{\rm RMS}$. In the bottom frame (b) we plot the binned density of the relaxed bodies (t=20h) and compare it to the model (dashed grey line). The Cartesian initial distributions clearly deviate from the model, expand and have a layer of low density material at the surface. Our initial conditions show an excellent match to the imprinted density profile and have no such low density ``atmosphere''.}}
	\label{fig:relaxedmodels}
\end{figure}

\comment{
In order to asses to performance of \texttt{ballic}, we compare it to two particle distributions of the same model, that are obtained using more conventional methods (Figure~\ref{fig:relaxedmodels}). Following previous work, we uniformly distributed the particles on a Cartesian grid inside of a sphere, so that a particle's density is equal to the uncompressed density, $\rho_0$, of the material. Of course such a uniform density is only physical for smaller bodies, as we expect larger bodies to compress under their own gravity resulting in a density gradient, and an Earth-mass body is thus poorly modelled.  We then imprint either a uniform \citep{Genda2012} or isentropic \citep{Canup2004} internal energy profile and let them evolve with our code until a self consistent equilibrium is reached. After 30h (in simulation time) we determine each particle representation's root mean square velocity, which measures the particle noise. As in previous work \citep{Canup2004}, \citep{Genda2012}, \citep{Marinova2011} we consider a model to be relaxed, if its root mean square velocity is below a small fraction of the impact velocity, in our case $v_{\rm RMS}$ < $50 ms^{-1}$. We also compare the relaxed body with the desired profile, to see how close the different methods are to the input model. The results are striking: the initial distribution obtained from \texttt{ballic} can be considered to be relaxed right from that start and even after 20h in simulation time still follows the imprinted density profile. The other two methods perform significantly worse both in terms of relaxation time, which is about 10h, and match to the original model. This is not surprising as these particle representations both suffer from Poison noise and start from unphysical initial conditions. We also experimented with re-imprinting an isentropic internal energy profile after every few steps of the simulation to enforce a closer match to the desired model and randomly distributing particles in concentric shells as in \citet{Marinova2011} but both proved to be difficult and time consuming as particles would enter unphysical states (see Section~\ref{section:isph} for details) which caused the hydro code to crash.}

\begin{figure}
	\centering
	\includegraphics{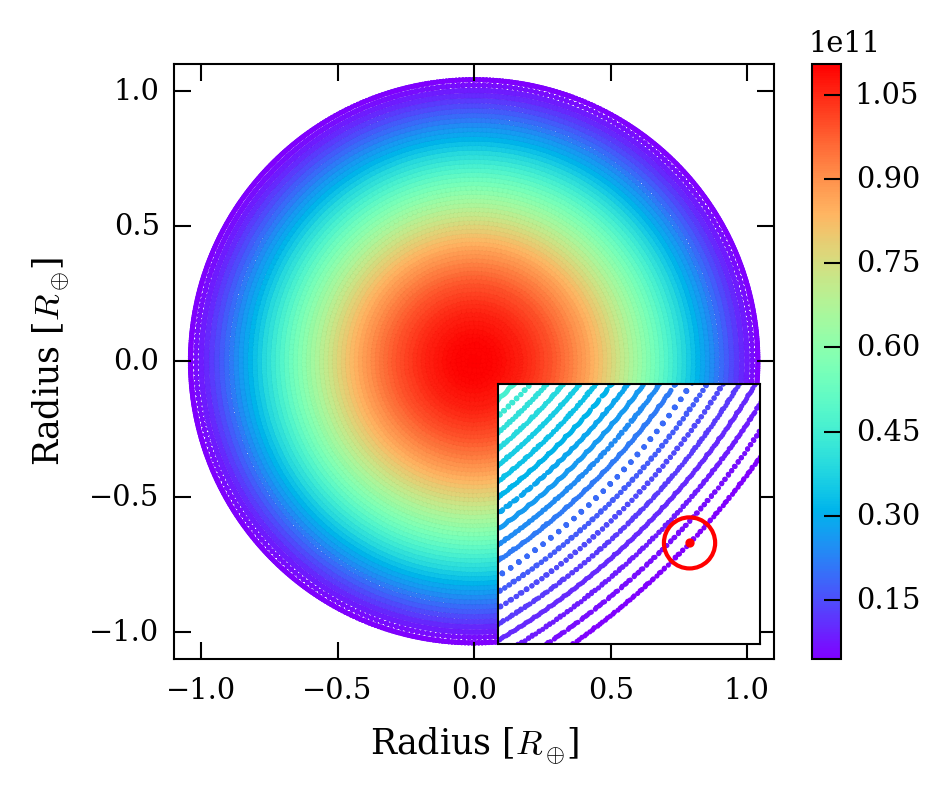}

	\caption{A slice (2.2 x 2.2 x 0.2 $R_{\oplus}$) through the SPH representation of an equilibrium model ($0.997 M_{\oplus}$ and $10^{5}$ particles) where the colours represent the internal energy of the material. The small plot shows a zoom at the surface of the model, where we marked the smoothing kernel (in red) of a particle in the outer most shell. Obviously such outer most particles have only one side of their kernel sampled by neighbouring particles while the rest is vacuum.}
	\label{fig:slice}
\end{figure}

\section{A modified density estimator for free surfaces}
\label{section:newdens}

In Smoothed Particle Hydrodynamics the density of a particle is derived from the masses and positions of its neighbouring particles \citep{Monaghan1992, Springel2010, Price2012}
\begin{equation}
\rho_i = \sum_{j=1}^{N}{ m_j W \left( \left| \vect{r}_i - \vect{r}_j \right| \right) }
\label{eqn:sphdensym}
\end{equation}
where the kernel function $W \left(r \right)$ weights each particle's contribution according to its distance from the central particle. Modern SPH implementations use a compact kernel and enforce a fixed number of nearest neighbour particles \citep{Monaghan1992} or enclosed mass \citep{Springel2002} to automatically adapt the resolution of the method. When the density is estimated in this way a particle at the planet's surface has about one half of its kernel sampled with neighbours while the rest of the volume is empty, or rather filled with vacuum. This means that the mass to volume ratio, and thus the resulting density, is underestimated because the standard SPH density estimator assumes that the whole kernel is (more or less homogeneously) filled with particles.

\begin{figure}
	\centering
	\includegraphics{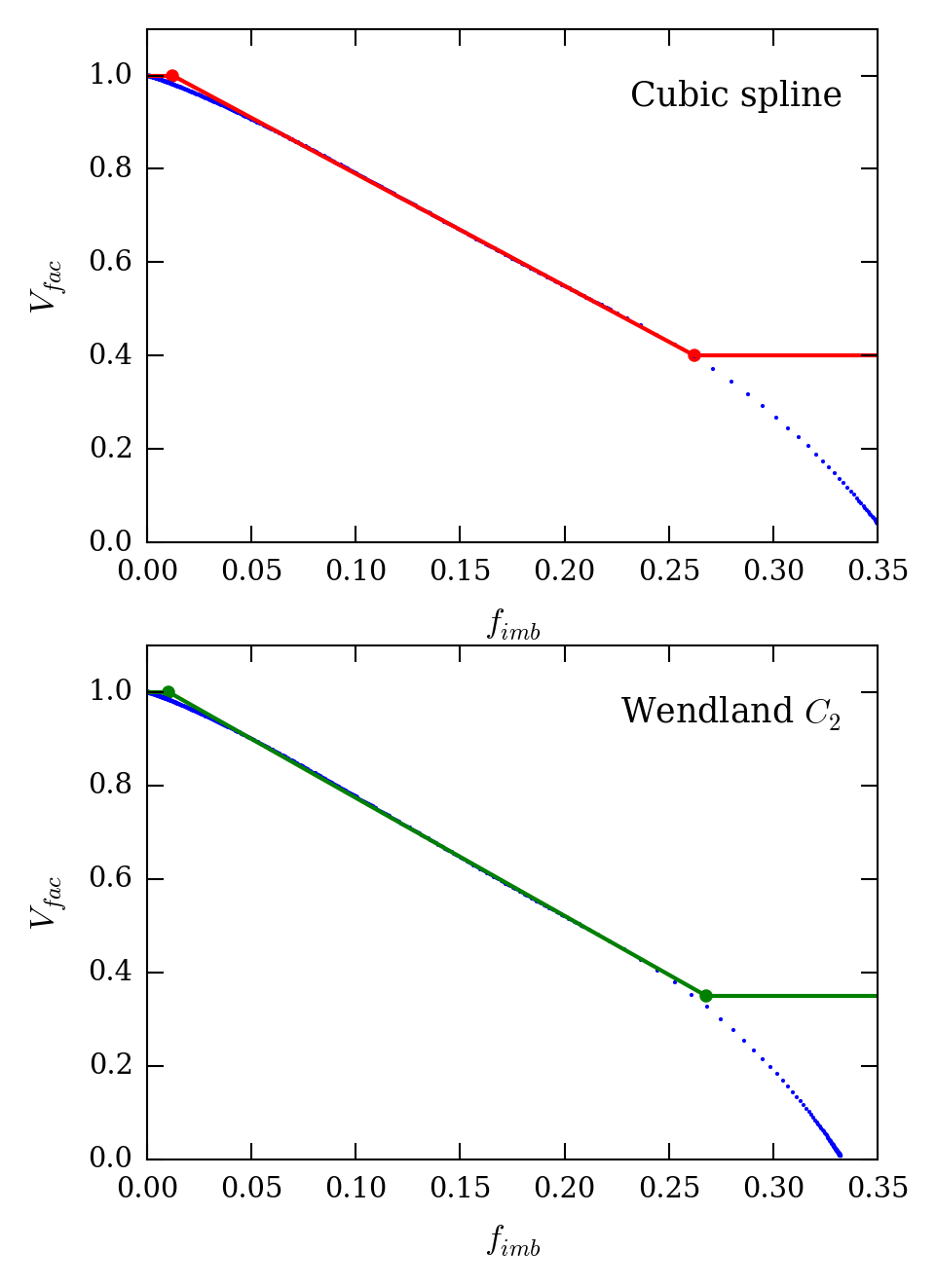}

	\caption{\comment{A plot that shows $V_{\rm fac} \left( f_{\rm imb} \right)$ for the cubic spline (top) and Wendland C2 kernel (bottom). Rather than using the obtained numerical result (blue dots) we fit a linear function (red and green lines) and keep the volume correction constant for small and large imbalances to avoid artificially triggered imbalance due to noise or large density gradients. The coefficients we obtained for the linear fit are $a=-2.39701$, $b=1.02918$ for the cubic spline and $a=-2.52444$, $b=1.0259$ for the Wendland C2 kernel.}}
	\label{fig:vfac}
\end{figure}

One possible treatment of vacuum/material interfaces is thus to correct the density using a correction factor
\begin{equation}
V_{\rm fac} = \frac{V}{V_{\rm eff}}
\end{equation} 
that only accounts for the effectively sampled volume $V_{\rm eff}$. One way to estimate $V_{\rm eff}$ is to consider the imbalance of the particle distribution in the kernel
\begin{equation}
f_{{\rm imb},i} = \frac{\left| \vect{f_i} \right|}{2 h_i \sum_j{ m_j W_{ij} \left( h_i \right)}}
\label{eqn:fimb}
\end{equation} 
where we used
\begin{equation}
\vect{f_i} = \sum_j{ \left( \vect{r_i} - \vect{r_j} \right) m_j W_{ij} \left( h_i \right)}
\label{eqn:fi}
\end{equation}
to measure the asymmetry of the particle distribution inside the kernel. We also tried to estimate the imbalance as suggested by Woolfson \citep{Woolfson2007} but found that the lower order of this estimator results in density fluctuations larger than the required correction. By using a kernel average (we use the Wendtland $C_2$, a fourth order kernel, with 80 nearest neighbours) as shown in the above equations our resulting density at the surface is much less noisy. The modified density is then obtained from
\begin{equation}
\rho_{\rm eff} = \frac{\rho_{\rm SPH}}{V_{\rm fac}}
\end{equation}
which obviously gives the standard SPH density if the particle distribution inside of the kernel is symmetric ($V_{\rm fac} = 1$). 
\comment{For a given filling factor of the kernel, where we assume a plane boundary between mass and vacuum, we can calculate the 
corresponding $f_{{\rm imb},i}$ by integrating Equation~\ref{eqn:fimb} numerically over the filled region. This is shown by the blue dots in figure \ref{fig:vfac}.}

Since the result was too sensitive to noise for very small and large imbalances we used a linear approximation (shown in green in Figure~\ref{fig:vfac}) and kept $V_{\rm fac}$ constant in these cases.

Since the proposed density correction only depends on a given particle's kernel volume, it is not symmetric (e.g., Equation~\ref{eqn:fimb} and \ref{eqn:fi}) and we have to calculate the standard SPH density from
\begin{equation}
\rho_i = \sum_{j=1}^{N}{ m_j W_{ij} \left( h_i \right)}
\end{equation}
\comment{which differs from that typically used in GASOLINE \citet{Wadsley2004}. It is not clear how to relate imbalance to a $V_{\rm fac}$ in a consistent way when using GASOLINE's symmetrized density, a formulation which improves the behaviour for very strong density discontinuities because a particle in the high density region still gets a contribution from low density particles. Correctly handling material interfaces, e.g., at core--mantle boundaries, is of greater relevance for GIs than this refinement to the density definition. Furthermore, we use a much larger number of neighbours, for which it is even less of an issue.} However, we do still use a symmetric formulation to evaluate the equation of motion such that both momentum and energy are conserved (e.g. \citet{Monaghan1992}, \citet{Wadsley2004}, \citet{Springel2010}, \citet{Price2012}).

Since the usually used B-spline kernels (see e.g. \citet{Wadsley2004} or \citet{Monaghan1992}) can trigger an instability that causes particles to clump when they are getting close we implemented the Wendland kernels proposed by \citet{Dehnen2012} that are both stable and allow a larger effective number of particles thus decreasing the noise for a given resolution. Another cause for clumping, independent of the chosen kernel function, is a negative pressure between particles. This is problematic as the Tillotson EOS calculates a negative pressure in the low density, cold region (Appendix~\ref{appendix:eos}) to mimic a tensional force between particles that are separated more than their equilibrium distance \citep{Melosh2007}. Since a fluid at these low densities (and temperatures) is expected to fragment and, for example, form droplets rather than behaving like a continuum we follow Melosh \citep{Melosh1989} and set the pressure to zero if it would become negative.

\begin{figure}
	\centering
	\includegraphics{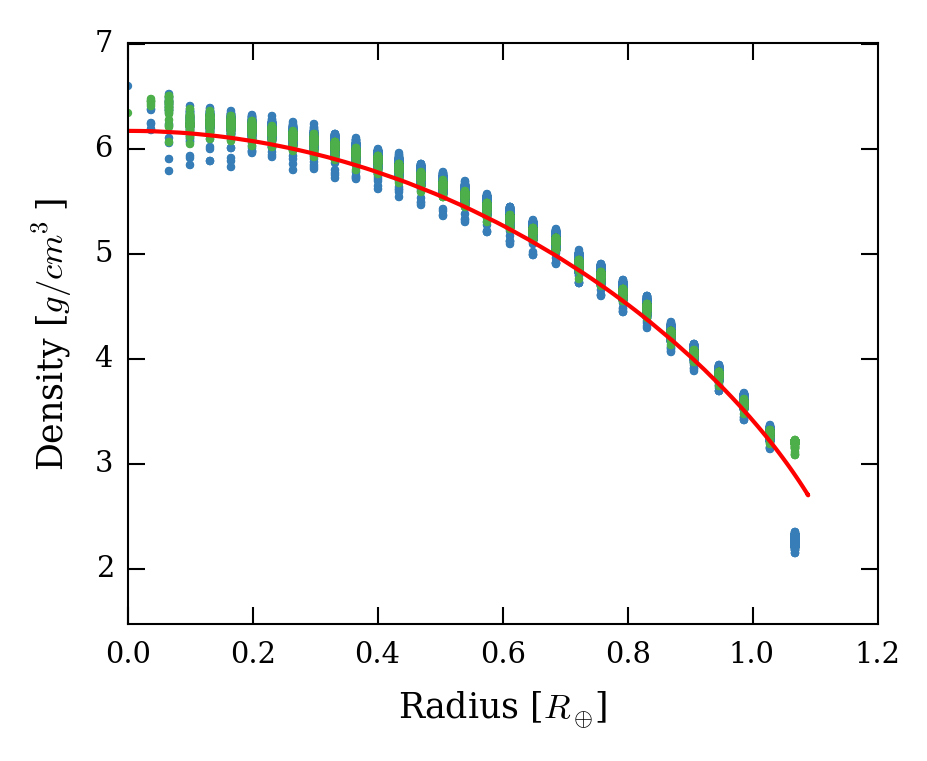}
	\caption{A density profile of the particle representation of an $M=0.997 M_{\oplus}$ protoplanet using $N=10^{5}$ particles. The solid red line shows $\rho(r)$ for the calculated equilibrium. The dots are the smoothed density for the classic SPH (light blue) and our new density estimator (light green). While the results agree for both methods in the inner part of the sphere, the particles in the last shell match the imposed surface density much better for our improved estimator. At this material/vacuum interface only a fraction of the kernel is sampled with SPH particles so we correct the standard SPH density accordingly. Besides avoiding an artificial "atmosphere" of low density material this also provides increased stability to the equilibrium models, so no relaxation is needed prior to use in a simulation. We also observe that our proposed modification tends to slightly overestimate the density compared to the desired values because imbalance of the particle distribution associated with the density gradient triggers an over--correction.}
	\label{fig:densityprofile}
\end{figure}

The resulting particle representations using the modified density estimator are in excellent agreement with the model (Figure~\ref{fig:densityprofile}) and the particles in the outer most shell remain in the condensed state. The smoothed density profile exhibits a general trend to over estimate the density compared to classic SPH, which is most pronounced in the outer most shell. This occurs because the proposed estimator artificially triggers an imbalance due to a density gradient on the kernel scale. For less steep density gradients, e.g., for lower mass models, the corrected density lies even closer to the desired values at the surface. In order to properly represent very massive planets, like super--earths, one would have to modify the procedure that relates $V_{\rm fac}$ to $f_{{\rm imb},i}$ to account for a density gradient. At this point we leave further improvements of the estimator to later work.

\section{Entropy conserving Smoothed Particles Hydrodynamics}
\label{section:isph}

\begin{figure}
	\centering
\begin{tikzpicture}[scale=0.6][cross/.style={path picture={%
		}}]%

	\draw [->, dashed, red, very thick] (1,1) -- (1,3);
	\node [right, shape=circle, draw] at (1.1,2) {1};
	\draw [->, dashed, red, very thick] (5,1) -- (5,3);
	\node [left, shape=circle, draw] at (4.9,2) {2};
	\draw [->, dashed, red, very thick] (1,3) -- (4,3);
	\node [above, shape=circle, draw] at (2,3.1) {3};

	\draw [->] (0,1) -- (6,1);
	\draw (0,5) -- (6,5);
	\node [right] at (6,1) {$\rho$};
	\draw [->] (1,0) -- (1,6);
	\draw (5,0) -- (5,6);
	\node [above] at (1,6) {$v$};
	\filldraw (1,1) circle (0.05);
	\node [below left] at (1,1) {$\left( u_{i,j}, \left. \frac{\partial u}{\partial \rho}\right|_{i,j} \right)$};
	\filldraw (5,1) circle (0.05);
	\node [below right] at (5,1) {$\left( u_{i+1,j}, \left. \frac{\partial u}{\partial \rho}\right|_{i+1,j} \right)$};
	\filldraw (5,5) circle (0.05);
	\node [above right] at (5,5) {$\left( u_{i+1,j+1}, \left. \frac{\partial u}{\partial \rho}\right|_{i+1,j+1} \right)$};
	\filldraw (1,5) circle (0.05);
	\node [above left] at (1,5) {$\left( u_{i,j+1}, \left. \frac{\partial u}{\partial \rho}\right|_{i,j+1} \right)$};

	\node [cross out, draw] at (1,3) {};
	\node [left] at (1,3) {$\left( u_{i,v}, \left. \frac{\partial u}{\partial \rho}\right|_{i,v} \right)$};

	\node [cross out, draw] at (5,3) {};
	\node [right] at (5,3) {$\left( u_{i+1,v}, \left. \frac{\partial u}{\partial \rho}\right|_{i+1,v} \right)$};

	\draw [dashed, red, thick] (1,3) -- (5,3);
	\node [below] at (3,3) {Hermite function};

	\filldraw (4,3) circle (0.05);
	\node [above] at (4,3) {$u \left( \rho, v \right)$};

\end{tikzpicture}

	\caption{A schematic explaining the 2D interpolation method used in ISPH. The values of $u(\rho,v)$ and $\left. \partial u/\partial \rho\right|_v$ are calculated from Equation~\ref{dudrho} while their derivatives with respect to v are obtained by fitting cubic splines in v. At each grid point $u(\rho,v)$, $\left. \partial u/\partial \rho\right|_v$, $\left. \partial^2 u/\partial v^2\right|_{\rho}$ and $\left. \partial^3 u/\partial \rho \partial v^2\right|_{\rho}$ are stored in the lookup table. The value of $u(\rho,v)$ is determined by (1) doing a spline interpolation in $v$ at $\rho_i$ to calculate $u_{i,v}$ and $\left. \partial u/\partial \rho\right|_{i,v}$, then (2) doing the same at $\rho_{i+1}$ to obtain $u_{i+1,v}$ and $\left. \partial u/\partial \rho\right|_{i+1,v}$. Step (3) is to interpolate $u \left( \rho, v \right)$ between these values using third order Hermite functions. Since $\left. \partial u/\partial v\right|_{\rho}$ and $\partial^2 u/\partial \rho \partial v$ can be determined from the second derivatives used for the cubic spline method $\left. \partial u/\partial v\right|_{\rho,v}$ can be calculated in the same lookup if needed.}
	\label{fig:splint}
\end{figure}
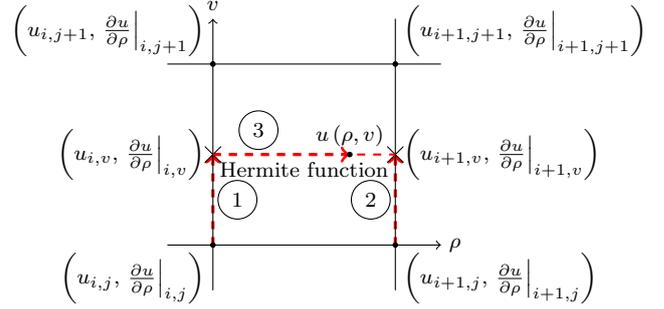

\begin{figure}
	\centering
	\includegraphics{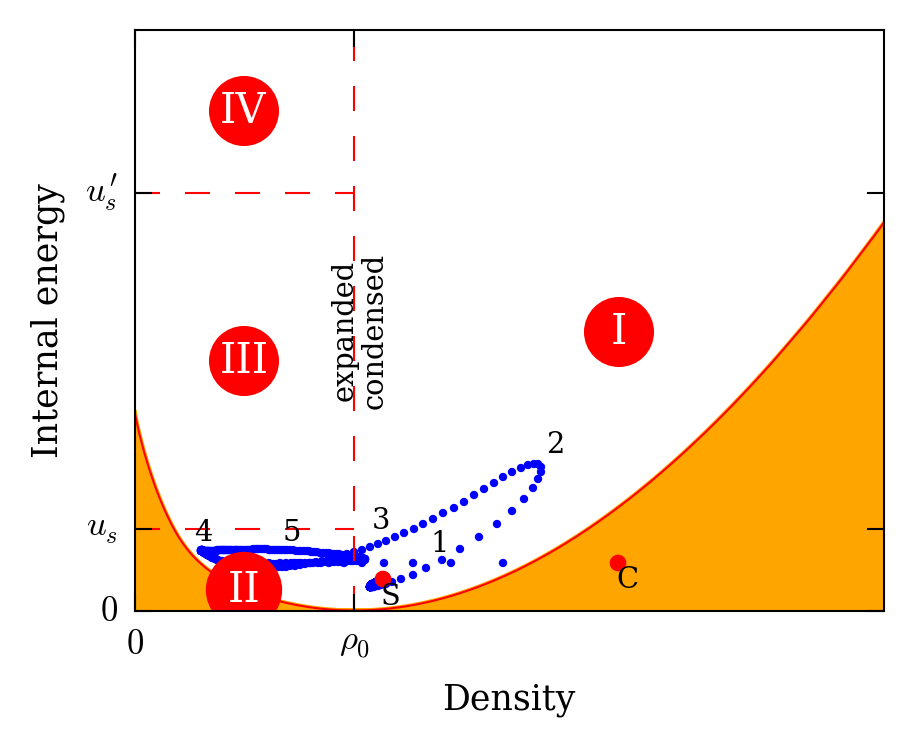}

	\caption{\comment{An example of a simulation ($b=0$, $v=10 km/s$, $N=10^5$) between an Earth-mass target and a Mars-sized impactor where the code crashed because a particle starting from (S) entered the unphysical region below the cold curve (C) of the Tillotson EOS. This happens because an ejected clump of material re impacts with the target and the particle's kernel suddenly overlaps with the target, which causes an unphysical increase in the density while the internal energy stays constant. If ISPH is applied to such a simulation, a particle can never evolve in such an way and the simulation does not crash. We also marked the four regions of the Tillotson EOS (I,II,III and IV) defined by the Tillotson parameters $\rho_0$,$u_s$ and $u^{\prime}_{s}$ as discussed in Appendix~\ref{appendix:eos}.}}
	\label{fig:crash}
\end{figure}

The results with the modified density were improved but some simulations would crash, e.g., due to unphysical entropy loss, even if the original ICs were setup in a physically consistent way as described above. For example, ejected material that re--impacts with the target would all of a sudden see a high density region when its kernel starts to overlap with the target, causing a sudden and unphysical increase in the particle's density while the internal energy stays constant (Figure~\ref{fig:crash}). This would not happen, if the method would exactly conserve the particle's entropy since its internal energy would increase accordingly.

Clearly evolving the entropy rather than the internal energy as a fundamental thermodynamical variable would be desirable. A particle will follow an isentrope for a purely adiabatic flow thus making it impossible to ever fall below the cold curve (if we start from physical initial conditions) because the only possibility is to move to a higher isentrope as a result of shock entropy generation.

In its standard version, GASOLINE evolves the internal energy as \citep{Wadsley2004}
\begin{equation}
\frac{du_i}{dt} = \frac{du_{ad}}{dt} + \frac{du_{\Pi}}{dt} 
\end{equation}
where 
\begin{equation}
\frac{du_{ad}}{dt} = \frac{P_i}{\rho_i^2} \sum_{j=1}^{n}{m_j \vect{v}_{ij} \cdot \nabla_i W_{ij}} 
\end{equation}
is the work from adiabatic compression or expansion (PdV work) and
\begin{equation}
\frac{du_{\Pi}}{dt} = \frac{1}{2} \sum_{j=1}^{n}{m_j \Pi_{ij} \vect{v}_{ij} \nabla_i W_{ij}}
\label{eqn:duavdt}
\end{equation}
is the contribution due to irreversible ($ds > 0$) shock heating. The artificial viscosity term $\Pi_{ij}$ (AV) is non zero only for converging flows and captures shocks (e.g. \citet{Monaghan1992}, \citet{Wadsley2004}, \citet{Springel2010}, \citet{Price2012}). Following previous work (e.g. \citet{Canup2004}) we used $\alpha = 1.5$ and $\beta=2\alpha$ for the viscosity parameters.

In the case of an ideal gas it is possible to define an entropy function $A(s) = (\gamma - 1) u \rho^{1-\gamma}$ \citep{Springel2002} so that one can alternatively evolve the entropy rather than the internal energy given by
\begin{equation}
\frac{dA_i}{dt} = -\frac{\gamma-1}{\rho^{\gamma}} \mathcal{L} \left( \rho_i, u_i \right) + \frac{\gamma-1}{\rho_i^{\gamma-1}} \frac{du_{\Pi}}{dt},
\label{eqn:Adot}
\end{equation}
where $\mathcal{L}$ is a source term. This has the advantage that in the absence of dissipative processes the entropy is exactly conserved ($\mathcal{L}=0$). This is not the case if the internal energy is evolved \citep{Springel2002}. The main difficulty of this entropy formulation is, however, that it requires the EOS to be both analytic and thermodynamically complete in order to define an entropy function. This is often not the case for a more complex EOS like the Tillotson EOS that we use. It is possible however to calculate the isentropes for an arbitrary EOS from the first law of thermodynamics
\begin{equation}
du = \frac{P}{\rho^2} d\rho + T ds
\end{equation}
where $ds = 0$ for constant entropy. So one can solve
\begin{equation}
\frac{du}{d \rho} = \frac{P}{\rho^2}
\label{dudrho}
\end{equation}
numerically and store the results for different initial values (taken at $u \left( \rho = \rho_0 \right)$) in a lookup table. Since this does not require an analytic expression of the pressure but just a numerical value it can be done for any EOS, even if it is tabular. In the absence of entropy changing processes, like shocks, the evolution of the internal energy for a given change in density can then be simply done by interpolating between these values.

To setup the lookup table we integrated Equation~\ref{dudrho} for different starting values $v_i = u_i \left( \rho = \rho_0 \right)$ in the direction of both increasing and decreasing densities. The variable $v_i$ thus can be used to label an isentrope and $v_0 = 0$ corresponds to the cold curve. For a given isentrope a particle's internal energy thus depends only on it's density $u = u \left( \rho, v \right)$. So in order to evolve a particle's energy from $\left( \rho_1, u_1 \right)$ to $\left( \rho_2, u_2 \right)$ we first have to determine on which isentrope it lies. For this we calculate the root of
\begin{equation}
F\left( v \right) = u_1 - u\left( \rho_1, v \right)
\end{equation}
which is bracketed by $v_0$ and $v_{max}$ using the Brent algorithm. To find $u \left( \rho, v \right)$ we do a two dimensional bi-cubic interpolation in the look-up table as described in Figure~\ref{fig:splint}. Since the derivatives of u with respect to $\rho$ are known analytically we only have to fit cubic splines along the direction of v (at constant density). Once a particle's isentrope is found we do the final interpolation of $u_2 = u_2 \left( \rho_2, v \right)$ for $\rho_2$.

To do realistic (impact) simulations one has to account for the change of entropy due to shocks thus we still need to account for the $d u_{\pi}/dt$ term. In SPH it is possible to obtain the internal energy contribution due to shocks from Equation~\ref{eqn:duavdt}. This allows one to split the (temporal) evolution of the internal energy into an adiabatic and a shock heating part, enforcing the conservation of entropy for an adiabatic flow and including the contribution of shock heating to the internal energy using the standard SPH formalism. \comment{This is very much in the spirit of a leapfrog--like evolution of the thermal energy and meshes well with the existing adaptive time--stepping leapfrog scheme used within GASOLINE.}

One time step from $t_i$ to $t_{i+1}$ in the \comment{ISPH adaptation of the} leapfrog algorithm used in GASOLINE\footnote{Here the Kick-Drift-Kick (KDK) version of the algorithm is described. This means that both the positions and forces are evaluated at the beginning and the end while the velocities are updated in the middle of the time step. This is convenient as the forces are evaluated at positions that are second order accurate.} is given by,
\comment{\begin{eqnarray*}
kick:		&	\vect{v}_{i+1/2} 	& = \vect{v}_{i} + \frac{1}{2} \vect{a}_{i} \Delta t\\
kick (entropy):	&	\vect{u}_{i}^{(1)}	& = \vect{u}_{i} + \frac{1}{2} \frac{du_{\Pi}}{dt} \Delta t\\
drift:		&	\vect{x}_{i+1}   	& = \vect{x}_{i} + \vect{v}_{i+1/2} \Delta t\\
drift (adiabatic):&	\vect{u}_{i}^{(2)}   	& = \vect{u}_{i}^{(1)} + \frac{du_{ad}}{dt} \Delta t\\
kick:		&	\vect{v}_{i+1}   	& = \vect{v}_{i+1/2} + \frac{1}{2} \vect{a}_{i+1} \Delta t\\
kick (entropy):	&	\vect{u}_{i+1}		& = \vect{u}_{i}^{(2)} + \frac{1}{2} \frac{du_{\Pi}}{dt} \Delta t
\end{eqnarray*}}
where $\vect{a}_{i} = \vect{a} \left ( \vect{x}_{i} \right)$ and $\vect{a}_{i+1} = \vect{a} \left ( \vect{x}_{i+1} \right)$ are the accelerations due to the forces acting between the particles.

Implementing such an integrator for a fluid is slightly more complicated because the forces not only depend on the position but also the velocity and internal energy of the particles \citep{Quinn1997}. These quantities are known at the start but not at the end of each time step so in order to calculate the fluid forces one has to use approximate predicted values to update the velocity from $v_{i+1/2}$ to $v_{i+1}$. 

We now describe a single ISPH time--step. 1) The gravitational and SPH forces on all particles $\vect{a} \left( t_i \right)$ and contributions from shock heating (Equation~\ref{eqn:duavdt}) are calculated at the beginning of the time interval. 2) Then the velocities and internal energies are updated for half a time step ("kick"). 3) Now all particle positions are "drifted" from $x \left( t_i \right)$ to $x \left( t_i + \Delta t \right)$. Before calculating the updated density $\rho \left( t_i + \Delta t \right)$ at the end of the interval its old value $\rho \left( t_i \right)$ is stored. 4) Using this value and the new density, calculate the change in internal energy due to adiabatic expansion or compression using the lookup table as described above. 5) In order to do the closing "kick" the \emph{predicted velocity} and \emph{predicted internal energy} at the end of the interval are determined. 6) The forces $\vect{a} \left( t_i + \Delta t\right)$ and shocking heating are calculated. 7) Finally both the velocity and internal energy are updated using the newly calculated force and shock heating term (closing "kick"). This way of evolving the energy equation matches very well with our time integration scheme and can be integrated rather easily into the code. The actual algorithm is more complex than this because the particles have their own unique time steps based on a block time stepping scheme. While particles are all drifted simultaneously on the smallest time step, such that the positions, and hence densities, are always time synchronized, those on larger time steps will not have their thermal energy advanced (see step 4 above) until they reach the end of their time step.

\begin{figure}
	\centering
\begin{tikzpicture}[cross/.style={path picture={
		}}]
		
	\draw [->] (0,0) -- (7,0);
	\node [right] at (7,0) {$\rho$};
	\draw [->] (0,0) -- (0,6);
	\node [above] at (0,6) {$u$};
		
	\filldraw (1,1) circle (0.1);
	\draw [dashed] (0,1) -- (1,1) -- (1,0);
	\node [left] at (0,1) {$u_{i}$};
	\node [below] at (1,0) {$\rho_{i}$};

	\filldraw (6,5) circle (0.1);
	\draw [dashed] (0,5) -- (6,5) -- (6,0);
	\node [left] at (0,5) {$u_{i+1}$};
	\node [below] at (6,0) {$\rho_{i+1}$};

	\draw [thick, red, ->] (1,1) -- (1,1.5);
	\node [left, red] at (1,1.25) {$\Delta u_{\pi}$};
	\draw [thick, red, ->] (1,1.5) parabola (6,4.5);
	\node [below, red] at (3.5,3) {$\Delta u_{ad}$};
	\draw [thick, red, ->] (6,4.5) -- (6,5);
	\node [left, red] at (6,4.75) {$\Delta u_{\pi}$};


	\draw [thick, blue, dashed, ->] (1,1) parabola (6,5);
	\node [below, blue] at (4.5,2.5) {$\Delta u_{num}$};
\end{tikzpicture}

	\caption{The evolution of the internal energy during one (sub) step in ISPH. The dashed blue line $\Delta u_{num}$ represents the exact numerical solution of evolution of the the internal energy as a weak shock occurs. ISPH divides the total change in $u$ into an adiabatic part $\Delta u_{ad}$ and shock heating $\Delta u_{\pi}$.}
	\label{fig:isph_u_evo}
\end{figure}
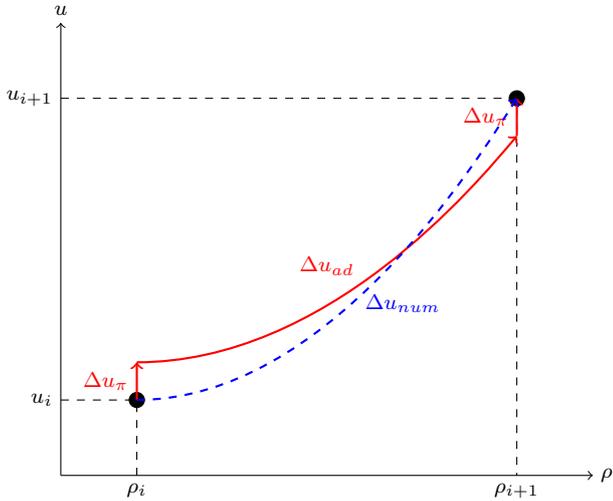

\section{Tests}
\label{section:tests}

As a first test we simulated the equilibrium models with our code to check if they truly stayed in equilibrium or would expand or collapse after a certain time. For classic SPH the models would remain stable but oscillate, as there was some intrinsic noise in our model and the density at the surface of the body was significantly underestimated as discussed in Section~\ref{section:newdens}. After several oscillations the particles would settle into an equilibrium configuration because the artificial viscosity in SPH would dampen their motion since it does not distinguish between a converging flow and a shock. As in previous studies (e.g. \citet{Canup2001}, \citet{Genda2012}) we used a velocity damper that reduced the particle's velocity each step to speed things up but this has to be done carefully to not over--damp the system as the models will become unstable again as soon as the damping term is removed. For simulations that used our proposed density correction the models remain stable right from the start and do not change at all. Using ISPH does, as expected, not affect the result in any way. 

\begin{figure*}
	\centering
	\includegraphics[keepaspectratio,width=\textwidth]{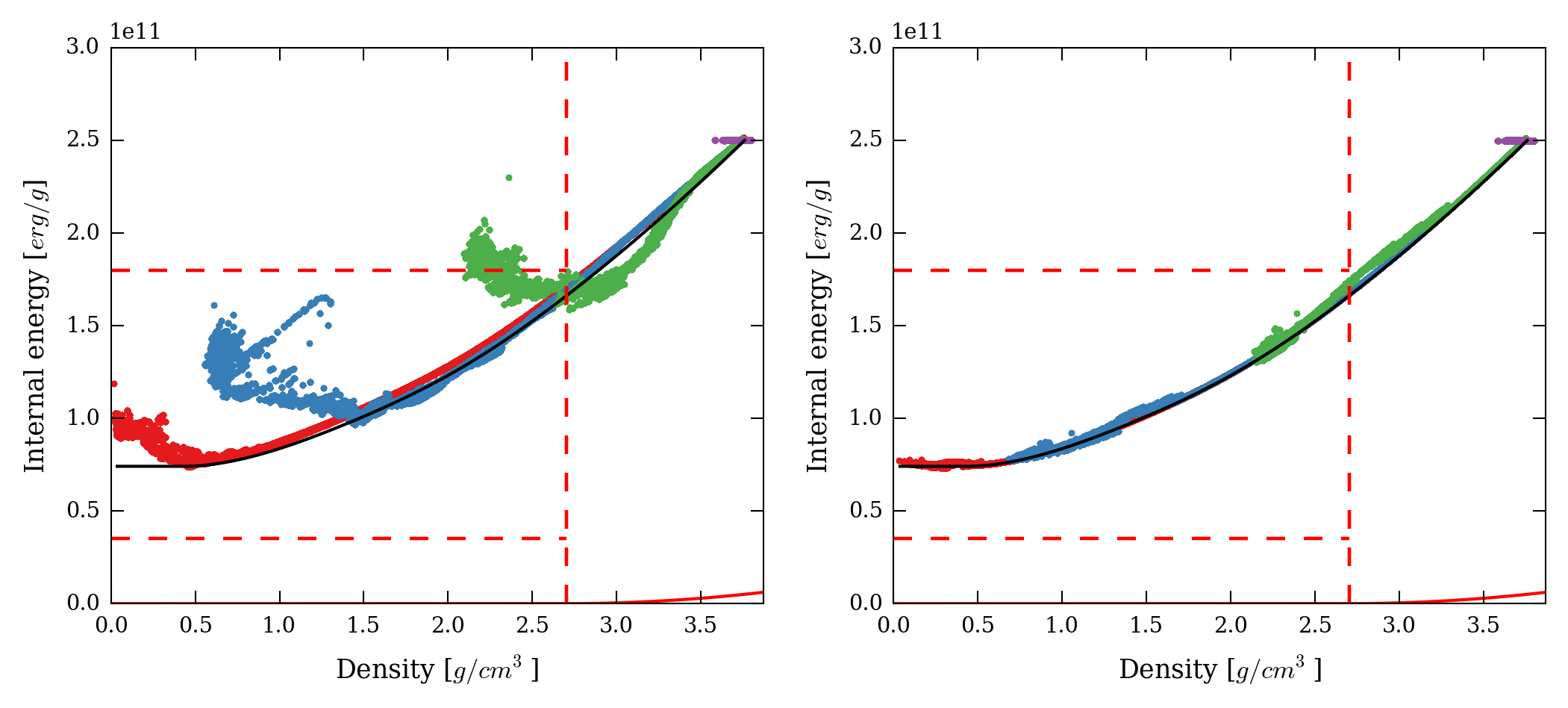}

	\caption{The adiabatic expansion of a uniform granite sphere. The left plot shows the results for classic SPH using our improved treatment of the material/vacuum boundary for different times (t=0s violet to \comment{t=0.7h} red) while the right plot does the same for ISPH. We clearly notice that for ISPH the particles follow their original isentrope, showing very little scatter in internal energy compared to classic SPH. Furthermore, all scatter that is observed in the ISPH case originates from the IC (the violet points at internal energy of $2.5 erg/g$).}
	\label{fig:adiabatic}
\end{figure*}

We then checked the code's ability to properly model an adiabatic flow which is not only a key requirement for any hydrodynamics code but also most important to test how accurately ISPH conserves entropy. For this we cut a sphere from a cubic uniform density and internal energy particle grid and let it evolve only due to pressure forces (neglecting gravity). We ran two simulations, one with classic SPH and one with ISPH both using our proposed modified density estimator. The differences are striking (Figure~\ref{fig:adiabatic}). While classic SPH suffers from huge scatter in the internal energy the particles in the ISPH simulation closely follow their initial isentrope over more than 17 hours in simulation time showing excellent entropy conservation.

Next we verified that we can recover the Rankine-Hugoniot jump conditions (e.g. \citet{Melosh1989}) relating the fluid quantities in front and behind the passing shock to test the code's ability to correctly capture shocks. For this we let two uniform granite slabs ($15 \times 15 \times 8 R_{\oplus}$) that are initially in contact collide with opposite velocities. The initial discontinuity in the velocity field then sends a shock wave through the material causing the fluid variables to jump to the post shock state at the shock front. These values can then be obtained from the simulations and compared to the theoretical results using the Rankine-Hugoniot equations. For both classic SPH and ISPH (with and without the modified density estimator) we obtain the correct values and thus conclude that shocks are adequately resolved in our code.

\begin{figure*}
	\centering
	\includegraphics[keepaspectratio,width=500pt]{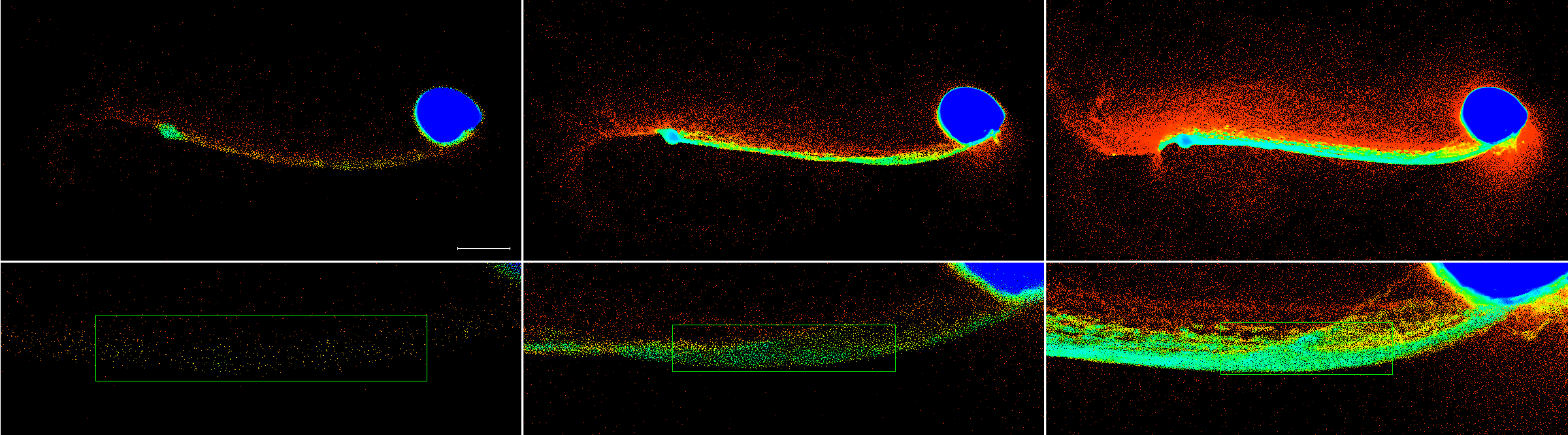}

	\caption{Simulation 1 is an example of an oblique ($b=0.71$, $v_{imp}=10 km/s$) collision between a $M=0.99 M_{\oplus}$ target and an $M=0.099 M_{\oplus}$ impactor (both bodies composed of granite only). The three top panels show the simulation 2.66h after the impact for different resolutions ($N=10^5$, $8 \cdot 10^5$, and $6.4 \cdot 10^6$ from left to right). The colours show the density from $0$ (deep red) to $2.7 g/cm^3$ (dark blue) corresponding to $\rho_{0}$ so dark blue material is condensed while everything else is in the expanded cold or intermediate states of the EOS. As in previous simulations on Moon forming impacts a part of the impactor survives the direct collision and forms a dense clump at the end of an "arm-like" structure. As shown in Figure~\ref{fig:ejecta} the impactor remnant's physical state is very similar for all resolutions and is composed only of about 1\% target material. The three bottom panels show a zoom into the above snapshots to show the "arm" that connects the target and the impactor remnant. For increasing number of particles the arm is much better resolved resulting in obvious differences in its structure and density profile (Figure~\ref{fig:ejecta}). A movie showing the early phase of the impact for the $6.4 \cdot 10^6$ particle simulation is available online (\url{https://youtu.be/QjpSg-gZl8U}).}
	\label{fig:grazingimpact}
\end{figure*}

\begin{figure}
	\centering
	\includegraphics{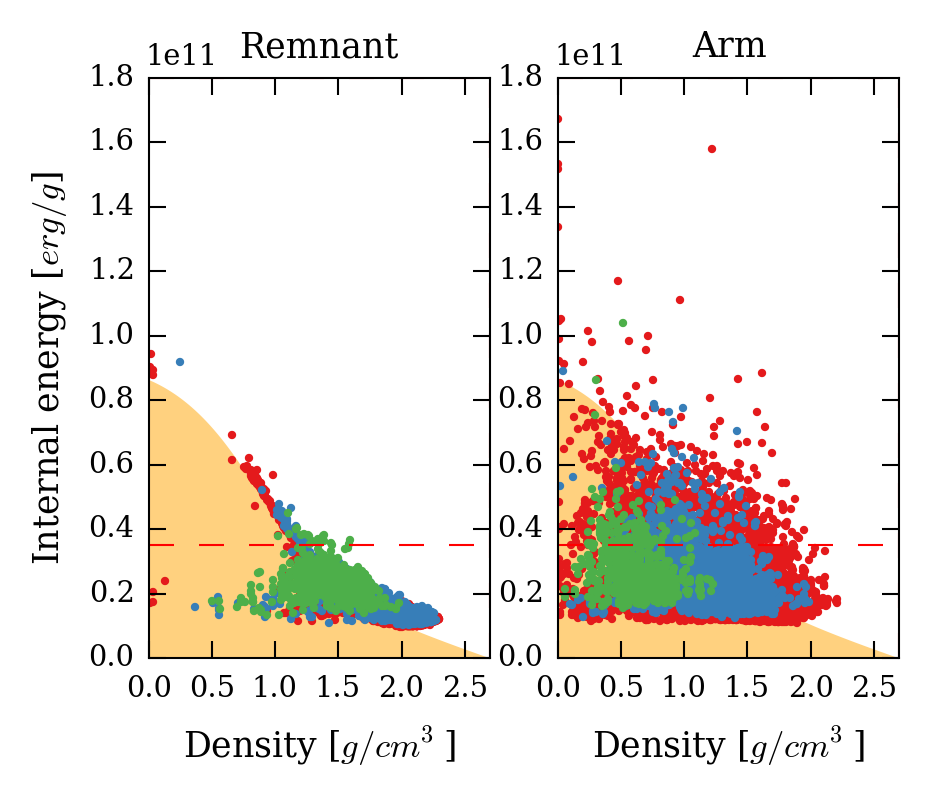}

	\caption{Results from a $b=0.71$, $v=10 km/s$ impact of a Mars-sized ($M=0.099 M_{\oplus}$) impactor onto an Earth-mass target ($M=0.99 M_{\oplus}$) \comment{shown in Figure~\ref{fig:grazingimpact}} for different resolutions of $10^5$ (green), $8 \cdot 10^5$ (blue) and $6.4 \cdot 10^6$ (red) particles. The top plot shows the internal energy vs density of the particles in the "arm" like structure that forms between the target and the surviving part of the impactor. For increasing resolution it becomes denser and more particles have an internal energy larger than $u_{IV}$ meaning that the material is partially evaporated. So the physical composition (and thus the pressure) in this low density region clearly depends on the resolution. The bottom plot shows the same for the impactor remnant that forms the end of the "arm" and consist mostly of impactor material that was not dispersed during the impact. The light orange region marks where we did the pressure cut--off because it would be negative for the Tillotson EOS. In this region the material is supposed to form small droplets and would ideally be treated by a multi--phase fluid prescription within the hydrodynamics.}
	\label{fig:ejecta} 
\end{figure}

As a last test we simulated collisions between granite spheres for different impact parameters and compared our results to prior work done in the field. Since there are not many publications on GI between undifferentiated bodies and the early work \citep{Benz1986} suffers from (for today's standards) low resolution and short simulation time, a direct comparison of the results was complicated. However we expect the basic features of such a collision to be similar for differentiated and undifferentiated bodies. First we run a simulation of an oblique impact ($b = 0.71$, $v_{imp} = 10km/s$, $M_t = 0.997 M_{\oplus}$ and $M_i = 0.099 M_{\oplus}$) like the one that might have created our moon \citep{Canup2001, Canup2004} using a total number of $10^5$ particles for both classic SPH without any of the modifications presented in this paper and ISPH (including the density correction discussed in Section~\ref{section:newdens}). Prior to the impact both bodies are placed in the xy-plane at a given separation so that they collide with the desired impact angle and velocity.
As in previous simulations the impactor is deformed due to tidal forces while approaching the target. The impact sends a strong shock through both bodies destroying most of the impactor and the ejected material forms an arm like structure for both methods which is consistent with prior results \citep{Canup2001}. Since our planetary models require no relaxation prior to the impact simulation we run the same simulation for an increasing number of particles for $N=10^5$,$8 \cdot 10^5$ up to $6.4 \cdot 10^6$ to check the resolution dependence of the results. While the general morphology (massive target, arm like structure and impactor remnant) is the same the density and internal energy of the material and thus the physical state of the material clearly depends on the resolution. This is most pronounced for the arm like structure as shown in Figure~\ref{fig:grazingimpact} and Figure~\ref{fig:ejecta}. This makes the use of a highly sophisticated EOS in low density, poorly resolved, regions questionable as the values of the fluid variables there are highly dependant on the resolution and the treatment of the hydrodynamics itself.

\begin{figure}
	\centering
	\includegraphics[keepaspectratio,width=0.46\textwidth]{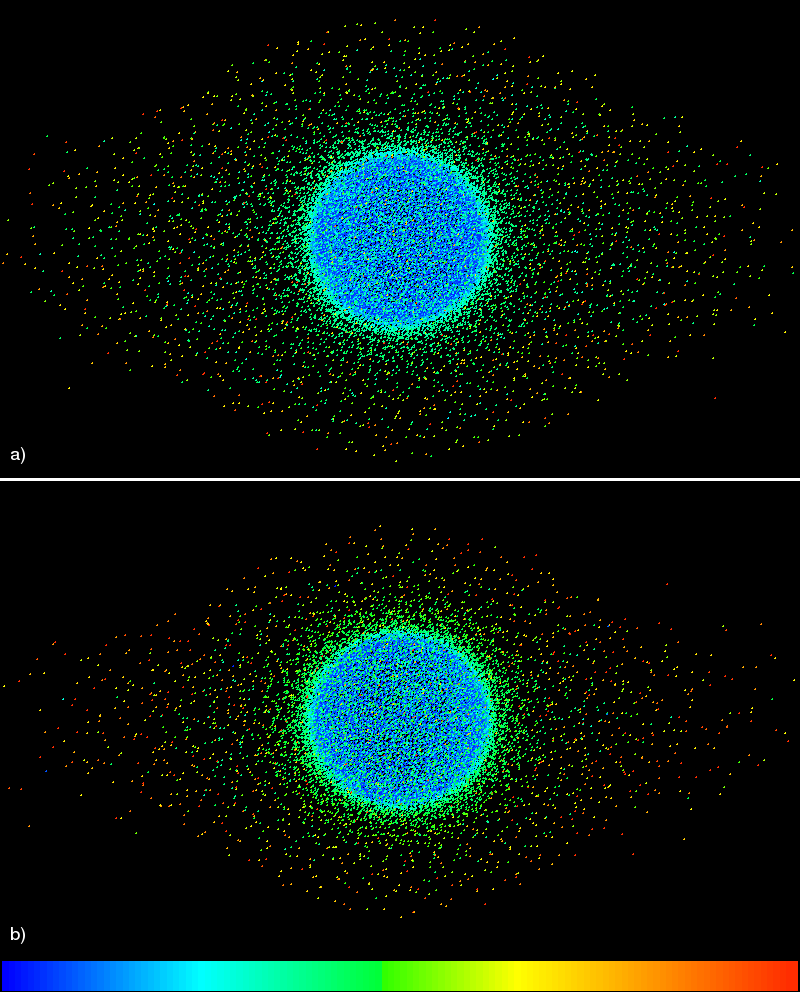}

	\caption{The circum--planetary disk resulting from a $b=0.5$, $v=10.61 km/s$ impact between a $M=0.9 M_{\oplus}$ target and an $M=0.2 M_{\oplus}$ impactor about three days after the collision. The colour shows the materials temperature from 4500K (blue) to 40'000K (red). The top frame (a) shows the result for classic SPH without any of the modifications presented in this paper and the bottom frame (b) shows the outcome when ISPH is used. For classic SPH the disk cooler, slightly more expanded and less flat.   
}
	\label{fig:disk_sph} 
\end{figure}

\begin{figure}
	\centering
	\includegraphics{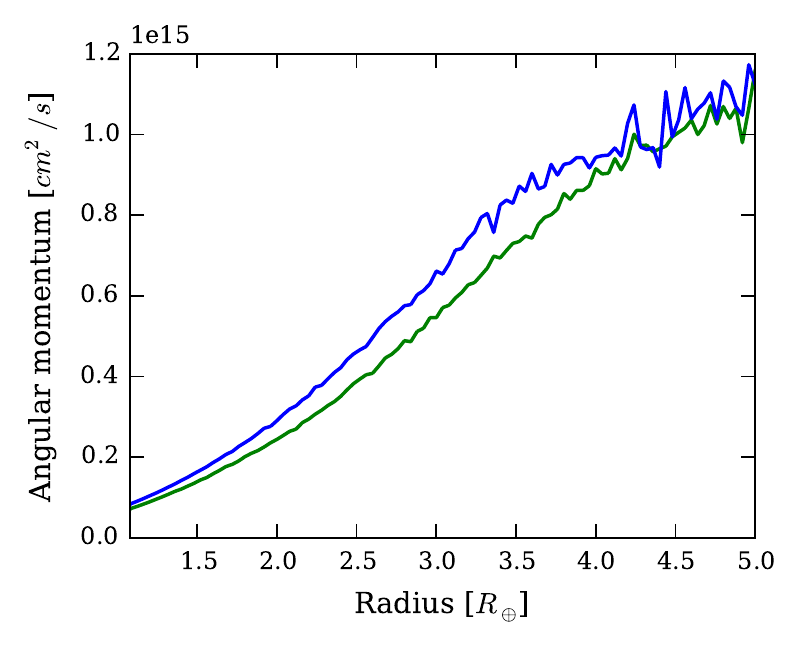}

	\caption{The specific angular momentum as a function of radius for the disk resulting from simulation 2 ($b=0.5$, $v=10.65 km/s$, $M_t=0.9 M_{\oplus}$ and $M_i=0.2 M_{\oplus}$) about three days after the impact as shown in Figure~\ref{fig:disk_sph}. The green line is result for classic SPH with none of the modifications discussed in this paper and the blue line is obtained using ISPH. In both cases we used the Balsara switch \citep{Balsara1995} to avoid artificial dissipation due to shear flows. While the methods agree within the first 24h after the collision the evolution is quite different at later times where the disk obtained using ISPH clearly has a larger angular momentum.}
	\label{fig:disk_plot}
\end{figure}

The second simulation we performed was a more head--on impact ($b=0.5$) of an $0.2 M_{\oplus}$ impactor onto a $0.9 M_{\oplus}$ target at $10.62$ km/s with $10^5$ SPH particles using classic SPH and ISPH. We let the simulation evolve for one day and again find that both methods initially produce very similar results. They do however differ in how the ejecta evolves on a longer time scale, as the impactor remnant leaves the system in the classic SPH simulation but remains bound when ISPH is used and re--impacts with the target after 4.4 days. In both cases the impact produces a disk of hot, partially vaporized material extending to about $3 R_{\oplus}$ from the target. Already after one day it assumes a flattened, rotationally supported profile which is more pronounced in the case of ISPH. We also observe a steep density gradient at the interface between the planet and the disk. To analyse disk properties we define it as the bound orbiting material that has a density lower that the reference density $\rho_0$ at the surface of the planet. Despite visible differences, for ISPH we obtain a slightly more massive planet and less massive disk than for classic SPH, the two methods produce very similar disk profiles up to 24 hours in simulation time. 
To study the long term evolution of the disk and the fate of the impactor remnant in the ISPH collision we continued both simulations for additional seven days beyond impact. As in previous simulations the Balsara switch \citep{Balsara1995} was used to reduce unwanted dissipation in the disk.
After 2.5 days the differences in the disk's structure due to the method are significant. The disk obtained from ISPH is rotating significantly faster (Figure~\ref{fig:disk_plot}) and is less extended than the one resulting in the classic SPH simulation. We also observe that for ISPH the disk remains flatter and is slightly hotter (Figure~\ref{fig:disk_sph}). These differences remain until the simulations ends eight days after the impact. At this point we leave a detailed investigation to later work.

\section{Summary and conclusion}
\label{section:conclusions}
In this paper we present three solutions to long standing problems of GI simulations with the SPH method. First we introduce a method to generate low noise particle representation of planets. Comparing the particle's (smoothed) density to the numerical solution of the equilibrium model shows excellent agreement and a low spread around the desired value. We then propose an improved treatment of a material/vacuum interface as it occurs at a planet's surface. There the classic SPH density estimator fails, calculating a significantly lower value. As a result the models obtained with \texttt{ballic} deviate from equilibrium and start to oscillate. This is even more problematic when the surface temperature is high as in this case the material is partially vaporized. As a solution we propose a modification of the standard SPH density estimator that accounts for only the part of the kernel volume that is sampled with particles, which gives a much better result and introduces less noise compared to previous work. Applying this density correction to our equilibrium models results in stable initial conditions that do not have to be relaxed prior to the simulation, significantly reducing the total simulation time allowing us to reach so far unprecedented number of particles. Finally we present an entropy conserving formulation of SPH that does not rely on an analytic or thermodynamically complete EOS. The evolution of the internal energy is split into a contribution from shock heating using the standard SPH artificial viscosity and an exact adiabatic evolution that is calculated by interpolating between isentropes. For standard SPH the particles significantly deviate from their original isentrope. Our new method conserves entropy and the particles never fall below their adiabatic evolution curves. At the same time, the Rankine-Hugoniot jump conditions for the fluid variables at the discontinuity are successfully recovered, so our scheme properly captures shocks as expected. 

A comparison between the new method and classic SPH when applied to GI simulations shows that the methods mostly agree on the general outcome within the first 24 hours after the impact. In one simulation (Simulation 2 in Section~\ref{section:tests}) the impact produces a flattened, rotationally supported disk of hot, low density material around the target. While the total bound mass is very similar, ISPH produces a slightly more massive central body and less massive disk compared to classic SPH. The other disk properties are, however, almost identical. To study the long term evolution of the ejecta and the disk we continued the simulation until eight days after the impact and observe that the differences between the methods become more pronounced. We also re--ran one simulation (Simulation 1 in Section~\ref{section:tests}) for $10^5$, $8 \cdot 10^5$ and $6.4 \cdot 10^6$ particles to check if the results within the first few hours after the impact agree and found a clear resolution dependence of the physical state of the material of the ejecta (Figure~\ref{fig:ejecta}). This suggests that the use of a more sophisticated EOS should be considered with care as these low density regions are poorly resolved and the results are dominated by the numerical resolution and not physics.

\section*{Acknowledgements}
We thank Prasenjit Saha for helping with the implementation of an early version of the ISPH lookup algorithm. CR also wishes to express his gratitude to Hidekazu Tanaka and Hidenori Genda for inviting him to the University of Hokkaido and the Tokyo Institute of Technology in Japan in 2014, and we want to thank them for the valuable discussions that we had during their visits in Zurich. All the simulations were performed on the zBox4 CPU cluster at the Institute for Computational Science (University of Zurich). This work was supported by the SNF Grant in ``Computational Astrophysics'' (200020\_162930/1).

\begin{appendix}
\section{The equation of state}
\label{appendix:eos}
An equation of state (EOS) is usually defined as a relationship between the pressure, density and internal energy/temperature of a substance. For GI simulations the proper choice of the EOS plays a key role as it not only has to be able to correctly model the behaviour of the material over a wide range of pressures, densities and internal energies but also must capture the strong shocks occurring in both bodies due to the impact. For our work we used the Tillotson EOS \citep{Tillotson1962} that has been used in many prior simulations (e.g., \citet{Benz1986}, \citet{Canup2001}, \citet{Marinova2011}, \citet{Genda2012}, \citet{Jutzi2013}) and was especially developed to model hyper-velocity impacts. Depending on the density and internal energy of the material it can be divided into four different regions where the pressure is given by a different analytic expression (Figure~\ref{fig:crash}). Each material is defined by 10 material constants (Table~\ref{tab:materials}). Despite its simplicity (compared to purely tabular EOS, maybe with analytic fits in different parts) the results are in good agreement with measurements \citep{Benz1986, Brundage2013} and it's ability to properly reproduce the materials Hugoniot curve, thus accurately modelling shocks, is excellent \citep{Brundage2013}. 

\begin{table*}
	\centering

	\begin{tabular}{|l|l l l l l l|}
\hline
Parameter		& Granite			& Basalt			& Iron				& Ice				& Olivine			& Water \\
\hline
a				& 0.5				& 0.5				& 0.5				& 0.3				& 0.5				& 0.5\\
b				& 1.3				& 1.5				& 1.5				& 0.1				& 1.4				& 0.9\\
$u_0$ [erg/g]	&$1.6 \cdot 10^{11}$&$4.87\cdot 10^{12}$&$9.5 \cdot 10^{10}$&$1.0 \cdot 10^{11}$&$5.5 \cdot 10^{12}$&$2.0 \cdot 10^{10}$\\
$\rho_0$ [g]	& 2.7				& 2.7				& 7.86				& 0.917				& 3.5				& 1					\\
A				&$1.8 \cdot 10^{11}$&$2.67\cdot 10^{11}$&$1.28\cdot 10^{12}$&$9.47\cdot 10^{10}$&$1.31\cdot 10^{12}$&$2.0 \cdot 10^{11}$\\
B				&$1.8 \cdot 10^{11}$&$2.67\cdot 10^{11}$&$1.05\cdot 10^{12}$&$9.47\cdot 10^{10}$&$4.9 \cdot 10^{11}$&$1.0 \cdot 10^{11}$\\
$u_s$ [erg/g]	&$3.5 \cdot 10^{10}$&$4.72\cdot 10^{10}$&$1.42\cdot 10^{10}$&$7.73 \cdot 10^{9}$&$4.5 \cdot 10^{10}$&$4.0 \cdot 10^{9} $\\
$u_s'$ [erg/g]	&$1.8 \cdot 10^{11}$&$1.82\cdot 10^{11}$&$8.45\cdot 10^{10}$&$3.04\cdot 10^{10}$&$1.5 \cdot 10^{11}$&$2.0 \cdot 10^{10}$\\
$\alpha$		& 5					& 5					& 5					& 10				& 5					& 5					\\				
$\beta$			& 5					& 5					& 5					& 5					& 5					& 5					\\
Reference		& Benz 1986			& Benz 1999			& Benz 1987			& Benz 1999			& Marinova 2011		& Woolfson 2007		\\
$c_v$ [erg/g]	&$7.9 \cdot 10^{6}$	& -					&$4.49\cdot 10^{6}$ & -					& -					& -					\\
\hline
\end{tabular}
\caption{The Tillotson EOS parameters for different materials. Depending on the reference they can vary a little bit. The specific heat capacity $c_V$ is not a Tillotson parameter and has to be found in the literature.}

	\label{tab:materials}
\end{table*}

For \emph{condensed states} ($\rho \geq \rho_0$) in region I the pressure is given by
\begin{equation}
P_{I,II} = \left( a + \frac{b}{\frac{u}{u_0 \eta^2} + 1} \right) u \rho + A \mu + B \mu^2
\label{condensedandcoldstates}
\end{equation}
where $\eta = \rho / \rho_0$ is the compression and $\mu = \eta - 1$ is the strain. For very large compression (partial) ionization of the material is described by the Thomas-Fermi model\citep{Brundage2013}.

The expanded states, where the density is smaller than the reference density $\rho_0$, are again divided into three regions. In region II (\emph{expanded cold states}) the material density is low but its internal energy is smaller than the energy at incipient vaporization ($u < u_s$) so it is still a liquid or a solid and the pressure is the same as for the condensed states. For expanded but very cold material the pressure can become negative which does not happen in region I. This corresponds to a tension in the sold material that prevents it from expanding like a gas as low densities. A fluid on the other hand will fragment into small droplets at these low densities and can not be described as a continuum anymore \citep{Melosh1989}. For this reason we set the pressure to zero if it would become negative and let the material evolve due to gravity only (similar to \citet{Hosono2016}). This also prevents the sound speed in the material from becoming unphysical (imaginary, see Appendix \ref{appendix:soundspeed} for the calculation of the sound speed) and avoids a numerical instability in SPH that causes unphysical particle clumping and occurs for negative pressures \citep{Dehnen2012}. It also ensures that the sound speed (used to model artificial viscosity in shocks and for setting the time step) stays non-negative in all regions of the EOS. In the zero pressure region, where the fluid description breaks down, we artificially set the sound speed to a small minimal value, in which case GASOLINE uses a fixed maximum time step (the so called {\em base time step}) to evolve the given particle.

If the internal energy is larger than the energy needed for complete vaporization ($u > u_s'$) the material is in region IV (\emph{expanded hot state}) where the pressure is given by
\begin{multline}
P_{IV} = a u \rho + \left(\frac{b u \rho}{\frac{u}{u_0 \eta^2} + 1} + A \mu \exp{\left[-\alpha \left( \eta^{-1} - 1 \right)\right]} \right) \\
		 \exp{\left[-\beta \left( \eta^{-1} - 1 \right)^2\right]}
\end{multline}
and the substance is completely vaporized. For very small densities the second term cancels and we asymptotically approximate an ideal gas\footnote{For an ideal gas the pressure is $P = \left( \gamma -1 \right) u \rho$ where $\gamma = C_p/C_v$ is the adiabatic index and $C_p$ and $C_v$ are the heat capacity at constant pressure or volume. For granite, the material used for most simulations in this paper, $a=0.5$ and we approximate an ideal gas with $\gamma=1.5$.} with $\gamma = a + 1$. It is important to note that in order to vaporize, the material does not only have to be very hot but also have a low density, so the greatest amount of vapour is not generated during the impact but right after when the material at the impact site expands as mentioned in \citet{Benz1986}.

In between those two regions  ($u_s < u < u_s'$) are the intermediate states (region III), where the pressure is a linear interpolation
\begin{equation}
P_{III} = \frac{P_e \left( u - u_s \right) + P_c \left( u_s' - u \right)}{u_s' - u_s}
\label{eqn:intermediate}
\end{equation}
between a low density solid/liquid and a vapour phase. This simple mixing rule prevents pressure discontinuities in the intermediate region but of course does not model mixed phases (e.g. liquid/gas) or phase changes which is a key weakness of the Tillotson EOS \citep{Benz1989, Canup2004, Brundage2013}. Especially for Moon-forming impacts, predicting the right amount of vapour generated by the impact is crucial since pressure gradients might play an important role in placing material into orbit \citep{Canup2004Review}. Another short coming in the context of the present work is that the Tillotson EOS is not thermodynamically complete, meaning that a second equation, e.g., relating internal energy to temperature is missing. For this reason one can not define an entropy function similar to the case on an ideal gas thus being restricted to the evolution of the internal energy in a simulation. As we have seen this does not limit the use of an entropy conserving scheme within SPH.  

\section{Calculating the sound speed for the Tillotson equation of state}
\label{appendix:soundspeed}
For numerical hydrodynamics the sound speed is very important as it determines the largest possible time step for the simulation. For SPH the sound speed also enters in the AV (Equation~\ref{eqn:duavdt}) that determines how much kinetic energy is converted to thermal energy at a shock front (e.g. \citet{Wadsley2004}). In order to derive the sound speed for an arbitrary EOS one has to either linearise Euler's equations and solve the corresponding eigenvalue problem or use the general expression for the sound speed in an ideal fluid
\begin{equation}
c^2 = \left. \frac{\partial P}{\partial \rho} \right|_{s}
\label{eqn:soundspeed}
\end{equation}
as we did for the Tillotson EOS. In regions I and II this gives
\begin{eqnarray}
c_{I,II}^2 & =	& \left( \Gamma_{I,II} + 1 \right) \frac{P_{I,II}}{\rho} + \frac{A + B \left( \eta^2-1 \right)}{\rho}\nonumber\\
		&  &  + \frac{b}{\omega_0^2} \left(\omega_0 - 1 \right) \left( 2u - \frac{P_{I,II}}{\rho} \right)\\
\Gamma_c & = & A + \frac{b}{\omega_0}\\
\omega_0 & = & \frac{u}{u_0 \eta^2} + 1
\end{eqnarray}
while on region IV we obtained
\begin{eqnarray}
c_{IV}^2 & = & \left( \Gamma_{IV} + 1 \right) \frac{P_{IV}}{\rho} + \frac{A}{\rho_0} e^{-\left( \alpha z + \beta z^2 \right)}\nonumber\\
	 &   & \left( 1+\frac{\mu}{\eta^2} \left(\alpha + 2 \beta z - \eta \right)\right)+\frac{b \rho u }{\omega_0^2 \eta^2}\nonumber\\
	 &   & e^{- \beta z^2} \left( \frac{2 \beta z}{\rho_0} \omega_0 + \frac{1}{u_0 \rho}\left( \frac{P_{IV}}{\rho}-2u \right) \right)\\
\Gamma_{IV} & = & a + \frac{b}{\omega_0} e^{- \beta z^2}\\
z	 & = & \frac{1}{\eta} - 1
\end{eqnarray}
from (\ref{eqn:soundspeed}). In the intermediate states (region III) the sound speed is
\begin{equation}
c_{III}^2 = \frac{c_{I,II}^2 \left( u - u_s \right) + c_{IV}^2 \left( u_s' - u \right)}{u_s' - u_s}
\end{equation}
because the pressure there is a linear interpolation between the cold and hot expanded states (Equation~\ref{eqn:intermediate}). Unlike in the case of an ideal gas, this expression can be negative in the cold expanded states region of the Tillotson EOS which results in an imaginary sound speed. Since this only occurs when the pressure becomes negative, it does not affect our simulations since we enforce non--negative pressures as discussed in Appendix~\ref{appendix:eos}. In order to have a well defined time step we simply impose a minimal value for $c_s^2$ in this case.
\end{appendix}

\bibliography{literature}{}
\bibliographystyle{mnras}

\end{document}